\documentstyle[12pt,epsfig]{article}
\begin{document}

\begin{center}

{\Large THERMODYNAMIC METRICS}

\

\

{\Large AND BLACK HOLE PHYSICS}

\vspace{15mm}

{\large Jan E. \AA man$^*$}\footnote{ja@physto.se}

\

{\large Ingemar Bengtsson$^*$}\footnote{ibeng@fysik.su.se}

\

{\large Narit Pidokrajt$^{**}$}\footnote{pidokrajt@gmail.com}

\vspace{15mm}

*{\sl Stockholms Universitet, AlbaNova\\
Fysikum\\
SE-106 91 Stockholm, Sweden}

\

**{\sl Sj\"olins Gymnasium S\"odermalm\\
Sandbacksgatan 10\\
SE-116 21 Stockholm, Sweden}

\vspace{15mm}

{\bf Abstract:}

\end{center}

\noindent We give a brief survey of thermodynamic metrics, in particular the 
Hessian of the entropy function, and how they apply to 
black hole thermodynamics. We then provide a detailed discussion of the Gibbs 
surface of Kerr black holes. In particular we analyze its 
global properties, and extend it to take the entropy of 
the inner horizon into account. A brief discussion of Kerr-Newman black holes is 
included.

\newpage

{\bf 1. Introduction}

\

\noindent This Special Issue concerns geometry in thermodynamics, and there is a 
branch of thermodynamics arising out of geometry---namely black hole thermodynamics. 
The fundamental relations that describe black holes have a number of special 
features. Many of them can be traced back to the fact that---unlike the thermodynamic 
systems that dominate textbooks on thermodynamics \cite{Landau, Callen, Frank}---black holes 
cannot spontaneously divide themselves into subsystems. Mathematically, their entropies 
are not extensive functions, and hence the usual arguments connecting stability to the 
concavity of the entropy function do not apply. 

Here we will be concerned with the information geometry of black holes. We 
take the rather restrictive view that this means that we are to study the 
(negative of) the Hessian of their entropy functions, interpreted as a metric 
tensor. This geometry is also known as the Ruppeiner geometry \cite{Ruppeiner1}, 
and its physical 
relevance is well established for a wide class of thermodynamic systems such as 
fluid and spin systems, but not really---despite a considerable amount of work---for 
black holes. There is a dream however, which is that it may have something to say 
about the obvious question: what properties must a function have if it is to serve 
as an entropy function for black holes?  The equilibrium states form a Gibbs surface, 
and the question is how we can recognize that a 
Gibbs surface with a given geometry and global shape is a black hole Gibbs surface? 
We have no answer to offer, but at least we can look for clues. 

We will begin our paper with a brief review of the foundations of Ruppeiner's 
geometry (sections 2 and 3). This is followed by a brief comment on entropic substances 
(section 4). We will then bring up the points where the discussion 
must be changed for non-extensive systems (section 5). In section 6 we give an 
in-depth discussion of the Ruppeiner geometry of the Kerr family of black holes, 
with special emphasis on the global nature of the Gibbs surface. 
In section 7 we partially extend this discussion to the Kerr-Newman family. The new 
results are summarized in section 8. In the end 
we will not arrive at a definite physical interpretation of the Ruppeiner 
geometry of black holes, but we hope that we have made a contribution towards this 
goal. The Ruppeiner geometry does encode 
information about the second variation of the entropy function for some of the 
most intriguing thermodynamic systems known.

\

\noindent {\bf 2. Ruppeiner geometry}

\

\noindent The Ruppeiner metric is defined on the Gibbs surface of a thermodyamic system 
using nothing but the fundamental relation $S = S(X)$, where $S$ is the entropy and $X^i$ 
are the remaining extensive variables of the system including its energy \cite{Ruppeiner1}. 
The metric tensor is simply the negative of the Hessian matrix,  

\begin{equation} g_{ij} \equiv - \frac{\partial^2S}{\partial X^i \partial X^j} \equiv  
- S_{,ij} \ . \end{equation}

\noindent (Note the notation for partial derivatives introduced in the second step. We will use it 
freely from now on.) A slightly older definition uses the Hessian of the energy as a 
function of the extensive variables including the entropy. This is called 
the Weinhold metric \cite{Weinhold}. These definitions are mathematically 
natural if there is an affine structure defined on the space coordinatized by the arguments of 
the functions. The metric preserves its form under affine coordinate transformations. 
(The analogy to K\"ahler metrics will strike readers familiar with the latter. 
In that case it is a complex rather than an affine structure that enables us to define 
the metric in terms of a potential without bringing in any non-zero Christoffel symbols.) 
A theory of Hessian metrics has been developed within mathematical statistics 
\cite{Amari}. In the simplest case one considers the (the negative of) 
the Hessian of the Shannon entropy. The resulting metric is known as the Fisher-Rao metric, 
and has an interpretation in terms of the distinguishability of nearby distributions under a 
large number of samplings. In this case there are clear connections to thermodynamic 
geometry \cite{Brody}. 

The notation of thermodynamics was not designed to do differential 
geometry. In particular functions are given names depending on the physical interpretation 
of the values they take, rather than on their mathematical form. Let us denote the 
extensive variables, including the energy $U$, as $X^i = (U, N^a)$. We may pass to the 
energy representation, where the extensive variables are $Y^i = (S, N^a)$. Intensive 
variables ${\mu}_i$, such as temperature, pressure and chemical potential, are usually 
defined through 

\begin{equation} dU = {\mu}_idY^i = TdS + {\mu}_adN^a \ , \label{15a} 
\end{equation}

\noindent while the intensive variables in the entropy representation have 
no agreed names. The fundamental relation takes one of the forms 
$U = U(S,N^a)$ or $S = S(U,N^a)$. The Weinhold metric is 

\begin{equation} ds^2_W = U_{,ij}dY^i\otimes_SdY^j = d{\mu}_j\otimes_S
dY^j = dT\otimes_SdS + d{\mu}_a\otimes_SdN^a \ , \end{equation}

\noindent and the Ruppeiner metric is 

\begin{equation} ds^2 = - S_{,ij}dX^i\otimes_SdX^j = - d\left(\frac{1}{T}
\right)\otimes_SdU +d\left(\frac{{\mu}_a}{T}\right) \otimes_S dN^a 
\ . \end{equation}

\noindent These two metrics are conformally related \cite{Ihrig}, 

\begin{equation} ds^2 = \frac{1}{T}(dT\otimes_SdS + d{\mu}_a
\otimes_SdN^a) = \frac{1}{T}ds^2_W \ . \end{equation}

\noindent As long as the temperature $T$ is positive both metrics are positive definite 
whenever $S$ is a concave function. Once these metrics have been written down in their 
defining representations one can perform coordinate changes freely. If the state space is 
two dimensional they can be diagonalized by using $T$ as a coordinate. 

Extensive systems have the property that (for $\lambda > 0$) 

\begin{equation} \lambda S(X) = S(\lambda X) \ , \end{equation}

\noindent and similarly for the energy function. Euler's theorem on homogeneous 
functions can then be used together with the first law to derive the Gibbs-Duhem relation:   

\begin{equation} U = Y^i\mu_i \hspace{5mm} \Rightarrow \hspace{5mm} 
dU = \mu_idY^i + Y^id\mu_i \hspace{5mm} \Rightarrow \hspace{5mm} 
Y^id{\mu}_i = 0 \ . \end{equation}

\noindent Then the thermodynamic metrics will be degenerate. Indeed we see that 

\begin{equation} d{\mu}_i = g^W_{ij}dY^j \hspace{5mm} \Rightarrow 
\hspace{5mm} Y^ig^W_{ij}dY^j = 0 \hspace{5mm} \Rightarrow \hspace{5mm} 
Y^ig^W_{ij} = 0 \ . \end{equation}

\noindent Hence every metric conformal to the Weinhold metric has a null eigenvector. 
In order to obtain a positive definite metric we must keep 
one extensive variable fixed. This is also required by the physical 
setting of thermodynamic fluctuation theory, which is usually assumed 
to concern a (sub-)system at fixed volume but fluctuating energy. 

\

{\bf 3. Physical interpretation and the role of curvature}

\

\noindent The Ruppeiner geodetic distance can be interpreted as follows. 
Choose a path on the Gibbs surface. Divide it into $N$ steps, where $N$ is 
large. Start the system at one of the chosen points, and change its environment 
so that it equilibrates to the next point on the path. To fix ideas, think of changing 
the state of a cup of coffee by moving it through a succession of $N$ rooms at different 
temperatures. An elegant argument  \cite{Salamon} then shows that the 
total entropy change of the ``universe''---the rooms and the coffee, in our example---in 
the limit of large $N$, is bounded from below by 

\begin{equation} \Delta S^{\rm U}
\geq 
\frac{D^2}{2N} \ , 
\end{equation}

\noindent where $D$ is the Ruppeiner length of the path. It follows 
that the lowest possible bound on $\Delta S^{\rm U}$ is obtained when the system is 
driven along a geodetic path. The Weinhold geodetic distance has an interpretation 
along similar lines  \cite{Salamon}. This result is of interest in finite time 
thermodynamics \cite{Berry}, and in computer simulations of small systems \cite{Cooks}. 

Ruppeiner originally introduced his metric in the context of thermodynamic 
fluctuation theory. We recall that this was initiated by Einstein, who 
inverted Boltzmann's formula $S = \ln{\Omega}$ (in the equiprobable case). 
The entropy assumes a 
maximum at equilibrium, and we can expand to second order in the fluctuations 
(assumed to be small). This leads to a probability distribution 
for the fluctuations which, using the properties of Gaussian integrals, is 
shown to have the form \cite{Landau}

\begin{equation} p(X) = Ce^{S(X)} 
\approx Ce^{S(X_0) + \frac{1}{2}\frac{\partial^2S}{\partial X^i 
\partial X^j}{\Big |}_{X_0}\Delta X^i\Delta X^j } = 
\frac{\sqrt{\det{g}}}{(2{\pi})^{\frac{N}{2}}}
e^{-\frac{1}{2}g_{ij}\Delta X^i\Delta X^j} \ , \label{Einstein} \end{equation}

\noindent where $g_{ij}$ is the Ruppeiner metric (and $N$ is the number of variables). 
All functions are evaluated at equilibrium and both sides of 
the equation transform like a density, as it should be. In the normalization 
we assumed positive definiteness of the metric, which is equivalent to the concavity 
of the entropy function. In a typical application one extensive variable is kept fixed, 
and provides a scale. Thus, if the volume of the fluctuating subsystem is 
fixed, we should work with 

\begin{equation} S = Vs(u,n) \end{equation}

\noindent where $u = U/V$ etc. At least in the 
small, the Ruppeiner distance now acquires a meaning in terms of distinguishability 
against the background of the fluctuations. 

The Riemann curvature scalar plays a major role in Ruppeiner's theory 
\cite{Ruppeiner1,Ruppeiner3}. 
In the first place it determines the accuracy of the approximation made in eq. 
(\ref{Einstein}). This is related to one of the main geometric roles of the 
curvature scalar, to determine the growth in volume of a small ball with its radius. 
In any curved space the volume of a ball of geodesic radius $r$ is related to 
the volume of a ball of radius $r$ in flat space through

\begin{equation} \frac{\mbox{Vol}(\circ )}{\mbox{Vol}_0(\circ )} = 1 - 
\frac{r^2R(0)}{6(d+2)} + \dots \ , \end{equation}

\noindent where $\mbox{Vol}_0$ is the volume of a ball in flat space, $d$ is the dimension, 
and $R(0)$ is the curvature scalar at the centre of the ball. This universal formula holds to 
lowest order in the radius $r$. The sign of the curvature 
scalar therefore carries important information. Finally the Riemann tensor governs the 
way geodesics deviate from each other.  

Ruppeiner argues that attractive forces between the microscopic 
constituents give positive curvature, repulsive forces give negative curvature.  
The sign of the curvature also distinguishes between the Bose and Fermi ideal 
gases \cite{Mrugala}. This picture is supported by numerous studies of model systems, 
especially fluid systems with two dimensional Gibbs surfaces \cite{Des,Ruppeiner5}, and in 
particular their behaviour at phase transitions. The Ruppeiner and the Weinhold geometry 
of the ideal gas are flat, equally so if the volume or the particle number is kept fixed. 

\

{\bf 4. Entropic substances}

\

\noindent A small comment on when the Ruppeiner metric is flat may be useful. The ideal 
gas is the prime example of substance 
in which the only forces acting on the constituents are constraint forces. Other examples 
exist, notably---to a good approximation---rubber. On the macroscopic level such substances 
obey caloric equations of state of the form $U = U(T)$. So, consider an inflated rubber 
balloon at room temperature. One might think that its radius is obtained by minimizing 
its energy subject to some constraints, but this is clearly wrong since its thermodynamic 
energy is unaffected by changes in the volume of the ideal gas or in the tension of the 
rubber. What does determine the radius of the balloon is that it maximizes its entropy. 
For this reason such materials go under the name entropic substances \cite{Muller}. 

Is it perhaps so that entropic substances always have flat thermodynamical 
metrics? For the Ruppeiner metric on a two dimensional Gibbs surface the 
answer is ``yes''. To avoid confusion concerning 
the notation, let us denote the energy by $x$, the other variables in the fundamental 
relation by $y$, and the entropy function whose Hessian we are interested in 
by $\psi = \psi (x,y)$. The material will be entropic if its temperature 
$1/\psi_{,x}$ is a function of the energy $x$ only, that is to say if there 
exists a function $f$ such that 

\begin{equation} \psi_{,x} = f(x ) \ . \end{equation}

\noindent It follows immediately (using overdots to denote derivatives) that 

\begin{equation} \psi_{,xx} = \dot{f} \hspace{10mm} \psi_{,xxx} = \ddot{f} \hspace{10mm} 
\psi_{,xy} = \psi_{,xxy} = \psi_{,xxy} = 0
\ . \end{equation}

\noindent The Riemann tensor of a Hessian metric involves second and third derivatives 
of $\psi$ only, and in two dimensions these conditions are enough to ensure that the 
Riemann tensor vanishes (and also that the Ruppeiner metric is diagonal in these coordinates). 
But this does not have to be so in higher dimensions, so the general answer to our 
question is ``no''. 
 
The Weinhold metric can still be curved, also for two dimensional systems. To see this 
we denote the entropy by $x$, the other variable in the fundamental 
relation by $y$, and the energy function whose Hessian we are interested in 
by $\psi = \psi (x,y)$. The material will be entropic if there exists a 
function $f$ such that 

\begin{equation} \psi_{,x} = f(\psi ) \ . \end{equation}

\noindent After some calculation one finds, for the curvature scalar of a two dimensional 
system, that 

\begin{equation} R_W = \frac{1}{2g^2}\dot{f}\ddot{f}\left[ f^2\psi_{,yy}^2 - 
2f\dot{f}\psi_{,yy}\psi_{,y}^2 + \dot{f}^2\psi_{,y}^4\right] \ . \label{entropic} 
\end{equation}

\noindent This does not vanish in general, although it does vanish when 
$\ddot{f} = 0$, so that $T$ and $U$ are linearly related. This is the case for the ideal gas. 

There is more to a geometry than its local curvature. Globally the Gibbs surface of the ideal 
gas when equipped with the Ruppeiner metric is an infinite flat plane if $S = S(U,V)$. 
If $S = S(U,N)$ it is an 
infinite covering of the flat plane with one point (at $N = 0$) removed \cite{Nulton}. We 
are not aware of a physical interpretation of this difference (although the latter result 
can be understood from general considerations \cite{Aman2}). 

\

{\bf 5. The transition to black holes}

\

\noindent We now come to black holes. They are equilibrium states of Einstein's relativity 
theory, and can be regarded as thermodynamic systems with the total mass $M$ of the spacetimes 
containing them serving as the energy function and one quarter of the area 
of their event horizon serving as their entropy. This truly remarkable result 
makes dimensional sense only because of the introduction of Planck's constant $\hbar$ 
(which is seemingly foreign to the theory) into the equations \cite{Bardeen, Bekenstein, Hawking}. 
 
Black hole entropy functions depend on mechanically conserved control parameters such 
as mass $M$, angular momentum $J$, and electric charge $Q$, so an affine structure is 
there. However, these entropy functions are not concave (unless the black hole is 
confined to a cavity, as is effectively the case if we add 
a negative cosmological constant), and the Ruppeiner metric is no longer 
positive definite. Many of the preceding arguments have to be rethought in 
this new context. The density of states grows fast with energy, which means that the 
canonical ensemble misbehaves \cite{Hawking}. Moreover, specific heats can be negative, 
with no immediate consequences for stability. Negative specific heats do in fact occur 
in self-gravitating gases in the Newtonian regime too, as is well known among astronomers 
\cite{LyndenBell}. The analogy should perhaps not be pushed too far, but like black holes 
self-gravitating gases cannot spontaneously divide themselves into parts, and they are 
therefore non-extensive systems. Early 
arguments about phase transitions caused by diverging specific heats \cite{Davies} 
are contradicted by analyses based on Poincar\'e's linear series method \cite{Katz, Arcioni}. 
For the Kerr black holes they show that that no instability is present in the 
microcanonical ensemble, given that the Schwarzschild black hole is stable (except for 
an intriguing hint of instability in the extreme limit \cite{Arcioni}). The Ruppeiner 
metric does not come into the question of the dynamical stability of a black hole at all, 
since this is a question about variations at constant ADM mass, not within a family of 
black holes described by a Gibbs surface. It does have consequences for black brane solutions 
however \cite{Hollands}. 

The role of the curvature scalar is somewhat changed in Lorentzian spaces since 
there do not exist round balls there. Their role is taken by causal diamonds, consisting 
of the intersection of a past and a future lightcone whose apices are connected by a 
timelike geodesic. The volume of a small causal diamond is \cite{Myrheim,Gibbons} 
(choosing our conventions so that de Sitter space has $R > 0$) 

\begin{equation} \frac{\mbox{Vol}(\diamond )}{\mbox{Vol}_0(\diamond )} = 
1 + \frac{d}{24(d+1)}\left( R_{ab}+\frac{R}{d+2}g_{ab}\right) t^at^b
+ \dots  \ , \label{myrheim} \end{equation}

\noindent where the right hand side is evaluated at the midpoint of the geodetic segment 
and $t^a$ is its tangent vector, normalized so that 

\begin{equation} g_{ab}(0)t^at^b = - \tau^2 \ . \end{equation}

\noindent A directional dependence has entered the formula. In two dimensions, where 
$R_{ab} = Rg_{ab}/2$, it enters only at the next order:

\begin{equation} \frac{\mbox{Vol}(\diamond )_2}{\mbox{Vol}_0(\diamond )_2} 
=  1 - 
\frac{R}{48}\tau^2 -\frac{1}{5760} 
\left( 4g_{ab}R^2 -g_{ab}\nabla^2R + 3\nabla_a\nabla_bR\right) t^at^b\tau^2  
+ \dots \ .  \end{equation}

\noindent 
It should also be noted that negative curvature, say, will cause timelike 
geodesics to converge and spacelike geodesics to diverge. 

The thermodynamic geometry of black holes was first discussed (briefly) by Page \cite{Page}, 
then by Ferrara et al. \cite{Ferrara}, and later by many others. In fact the 
literature is by now large and varied, as can be gleaned from the reference lists in some 
recent papers \cite{Ruppeiner5,Medved,Dolan}. As an example of a direction that we 
will not look into here, let us mention the idea that the non-equivalence of ensembles 
in the black hole case means that the Hessian of other thermodynamic potentials, such 
as the free energy, should be studied; see ref. \cite{Bravetti} and references therein. 

For our present purposes it must be mentioned that the scale symmetry of Einstein-Maxwell equations, 
with the cosmological constant set to zero but with the spacetime dimension $d \geq 4$ kept 
arbitrary, implies that the entropy of any black hole is a generalized homogeneous function. 
With $J$ being the angular momentum and $Q$ the electric charge there holds 

\begin{equation} \lambda^{d-2}S(M,Q,J) = S(\lambda^{d-3}M,\lambda^{d-3}Q, \lambda^{d-2}J) 
\ . \label{scaling} \end{equation}

\noindent It follows that the thermodynamic metrics will admit a homothetic Killing vector 
field, and also that there is a close relative of the Gibbs-Duhem relation. The 
intensive variables are the Hawking temperature $T$, the electric potential $\Phi$ at the 
horizon, and the angular velocity $\Omega$ of the horizon. The Hawking temperature is 
simply related to a geometrical property of the event horizon known as its surface gravity. 
From eq. (\ref{scaling}) one may deduce the Smarr formula \cite{Smarr}

\begin{equation} (d-3)M = (d-2)ST + (d-3)Q\Phi + (d-2)J\Omega \ . 
\end{equation}

\noindent For the Kerr and Reissner-Nordstr\"om families the implication is 
that there exist functions $f$ of one variable such that 

\begin{equation} S = M^{c_1}f(M^{c_2}X) \label{skala} \end{equation}

\noindent (where $X$ denotes either $J$ or $Q$), with exponents that are easily 
calculable. 
%
%
But every entropy function of the form (\ref{skala}) leads to a flat Ruppeiner metric 
if $c_2 = -1$, and a flat Weinhold metric if $c_1+c_2=0$ \cite{Aman2}. With no further 
assumptions about the form of the function $f$ it then follows that the Ruppeiner metric of 
the Reissner-Nordstr\"om families, and the Weinhold metric of the Kerr families, are both 
flat. 

In the next two sections we will go a bit beyond these known results, 
especially as concerns the global geometry of the state spaces. 


\noindent 

\

{\bf 6. Kerr black holes}

\

\noindent Given the interest in higher dimensional versions of physics, and the various 
matter fields suggested by all sorts of theories, there is a large and varied 
supply of black hole families to discuss. However, we will concentrate on 
the Kerr family, for the obvious reason that this is the only family that can make 
a solid case for actual existence. It is believed that many Kerr black holes do exist 
in the Milky Way, including a large one at its centre \cite{Kerr}. 

We need to know that the event horizon of a Kerr black hole with mass $M$ and 
angular momentum $J$ has area 

\begin{equation} A_+ = 8 \pi M^2\left( 1 + \sqrt{1 - J^2/M^4}\right) \ . \end{equation}

\noindent The event horizon exists only if the angular momentum is bounded by the 
inequality $-M^2 \leq J \leq M^2$ (in suitable units), which in everyday terms is a 
very strong constraint. The exact solution also has an inner horizon with area 

\begin{equation} A_- = 8 \pi M^2\left( 1 - \sqrt{1 - J^2/M^4}\right) \ . \end{equation}

\noindent If $J/M^2 = \pm 1$ the two horizons coincide and we have an extreme black 
hole with vanishing surface gravity (that is, vanishing Hawking temperature). Since the 
inner horizon will play an important role below we should perhaps say that there is no 
reason to believe that the spinning black hole at the centre of the Milky Way has an 
inner horizon. The sense in which that black hole is likely to be modelled by the Kerr 
solution is bound up with asymptotics, just as the equilibrium states and quasi-static 
processes of textbook thermodynamics are useful shorthands for a more complicated 
reality. 

We take the view that the inner horizon is an important feature of the equilibrium state. 
Its thermodynamics has already received some attention in the literature \cite{Curir,Kaburaki}. 
Consequently we have two distinct entropy functions to study, namely 

\begin{equation} S_{\pm} = S_{\pm}(M,J) = \frac{k}{4}A_{\pm} = 
2M^2\left( 1 \pm \sqrt{1-J^2/M^4}\right) \ . \end{equation}

\noindent (Following Davies we have set Boltzmann's constant $k = 1/\pi$ \cite{Davies}.) 
There will also be two different Hawking temperatures 

\begin{equation} T_\pm = \pm \frac{1}{4M}\frac{\sqrt{1-J^2/M^4}}{1 \pm \sqrt{1-J^2/M^4}} 
\ . \end{equation}

\noindent They both vanish in the extreme limit. Otherwise $T_+$ is positive and $T_-$ is 
negative. 

If we invert the entropy function to obtain the mass $M$ as a function of entropy and 
angular momentum we find the same functional form in both cases, 

\begin{equation} M = \sqrt{\frac{S_{\pm}}{4} + \frac{J^2}{S_{\pm}}} \ . \end{equation}

\noindent Consequently we will obtain the same expression for its Hessian, the Weinhold 
metric, in both cases---only the range of the coordinates will differ. 

The expressions for the Weinhold and Ruppeiner metrics in their defining coordinates 
are not very illuminating \cite{Aman1}. Changing to the dimensionless coordinate 

\begin{equation} a = \frac{J}{M^2} \ , \hspace{5mm} -1 \leq a \leq 1 \ , \end{equation}

\noindent and using our expression for $T_{\pm}$, we obtain for the two Ruppeiner metrics 

\begin{equation} ds^2_{\pm} = \frac{1}{T_{\pm}}\left[ - \frac{dM^2}{M} + 
\frac{M}{2}\frac{da^2}{(1-a^2)(1\pm \sqrt{1-a^2})} \right] \ . \end{equation}

\noindent The expression within brackets gives the Weinhold metric. To bring the latter 
to manifestly flat form we perform a sequence of coordinate transformations, viz. 

\begin{equation} a = \sin{2\beta} \ , \hspace{5mm} 
\cosh{2\alpha} = \frac{1}{\cos{\beta}} \ , 
\end{equation}

\begin{equation} t = 2\sqrt{M}\cosh{\alpha} \ , \hspace{5mm} x = 2\sqrt{M}\sinh{\alpha} 
\ . \end{equation}

\noindent The result is that 

\begin{equation} ds^2_{\pm} = \frac{1}{T_{\pm}}\left[  - dt^2 + dx^2\right] \ . \end{equation}

\noindent We must now consider the coordinate ranges. 

In the calculation we made use of the equation $\sqrt{1-a^2} = \cos{2\beta}$, so it is clear 
that the range $- \pi/4 \leq \beta \leq \pi/4$ must be used for the Ruppeiner metric $ds^2_+$ 
associated to the outer horizon. Its Gibbs surface is a timelike wedge, with a locally flat 
Minkowski metric for its Weinhold metric. The wedge is bounded by 

\begin{equation} - \sqrt{\frac{\sqrt{2}-1}{\sqrt{2}+1}} \leq \frac{x}{t} 
\leq \sqrt{\frac{\sqrt{2}-1}{\sqrt{2}+1}} = \tan{\frac{\pi}{8}} \ . \end{equation}

\noindent Its opening angle is $45^\circ$. The Ruppeiner metric itself is not defined on 
the edge of the wedge, since the conformal factor diverges there. However, the Weinhold 
metric can evidently be analytically extended. By increasing the coordinate range to 
$- \pi/2 \leq \beta \leq \pi/2$ we include also the Gibbs surface corresponding to 
the inner horizon. The combined Gibbs surface is isometric to the future 
{\bf null} cone of 
Minkowski space, as far as its Weinhold metric is concerned. We find this satisfying. 

Using the Minkowski space coordinates we can now give unifying expressions for the 
thermodynamic functions. The mass and the entropy are 

\begin{equation} M = \frac{t^2-x^2}{4} \end{equation}

\begin{equation} S = 2M^2(1 \pm \sqrt{1-a^2}) = 
M^2(1+\cos{2\beta}) = \frac{(t^2-x^2)^4}{4(t^2+x^2)^2} \ . \end{equation}

\noindent Both of them vanish on the light cone (while they remain finite on the edge 
of the wedge, where the extreme black holes sit). The Hawking temperatures $T_{\pm}$ 
are unified to 

\begin{equation} T = 
\frac{(t^2-x^2-2tx)(t^2-x^2+2tx)}{2(t^2-x^2)^3} \ . \end{equation}

\begin{figure}
\centering
\begin{minipage}[t]{0.45\linewidth}
\centering
\includegraphics[scale=.4]{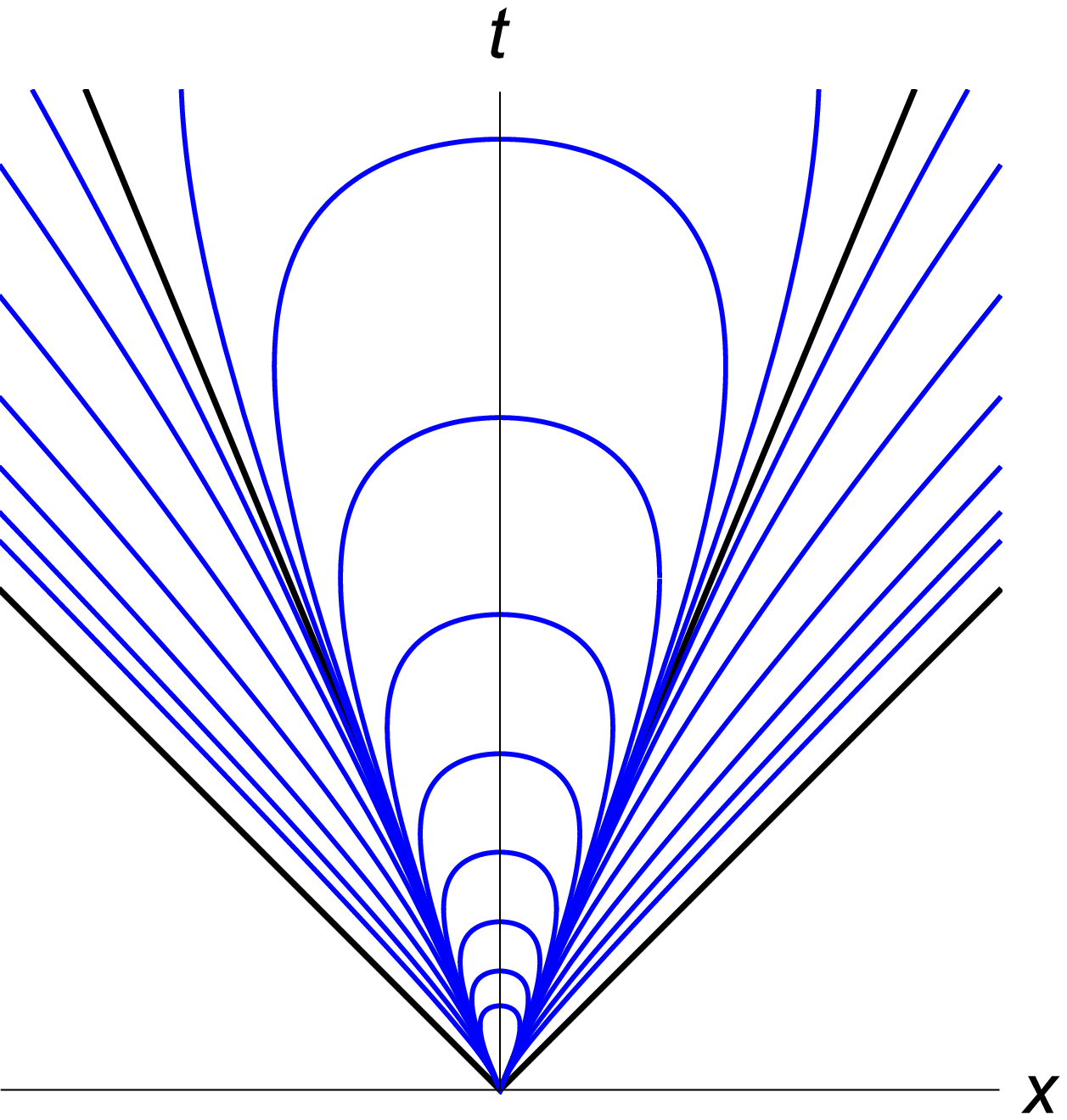}
\label{fig:Jan3}
\caption{\small  Contour curves of equal Hawking temperature $T$. The Hawking 
        temperature vanishes at the edge of the wedge that corresponds 
        to the outer horizon, is negative outside, and diverges on the 
			 null cone.}
\end{minipage} 
      \vspace{0pt} \hspace{.1cm}  
\begin{minipage}[t]{0.45\linewidth}
\centering
\includegraphics[scale=.4]{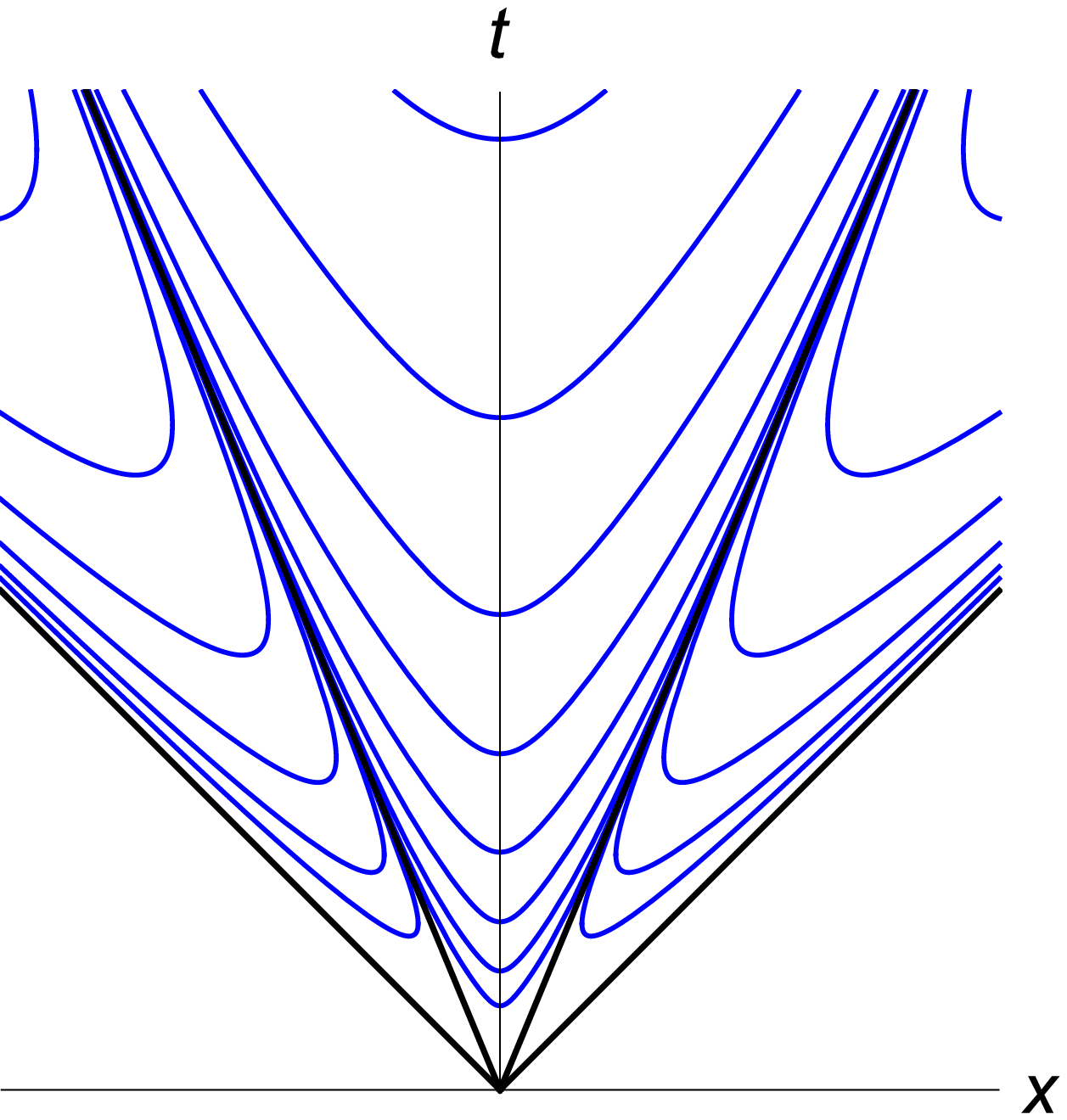}  
\caption{\small \noindent Contour curves of equal Ruppeiner scalar curvature $R$. 
It is negative inside the wedge, positive outside, and diverges both at the edge of 
the wedge and on the 
null cone.}   
\label{fig:Jan2}  
\end{minipage} 
\end{figure}

\noindent Its variation over the Gibbs surface is shown in Fig. \ref{fig:Jan3}.

The state space volume, as measured by the Ruppeiner metric, is strongly concentrated 
to the neighbourhood of the extreme black holes. The Ruppeiner curvature scalar is 

\begin{equation} R = 
- \frac{4(t^4+10t^2x^2+x^4)}{(t^2-x^2)^2(t^4-6t^2x^2+x^4)} \ . \end{equation} 

\noindent It diverges at the edge of the wedge and on the 
null cone, as seen in Fig. 
\ref{fig:Jan2}. 

\vspace{.cm}

\begin{figure}
\centering
\begin{minipage}[t]{0.45\linewidth}
\centering
\includegraphics[angle=90, scale=.4]{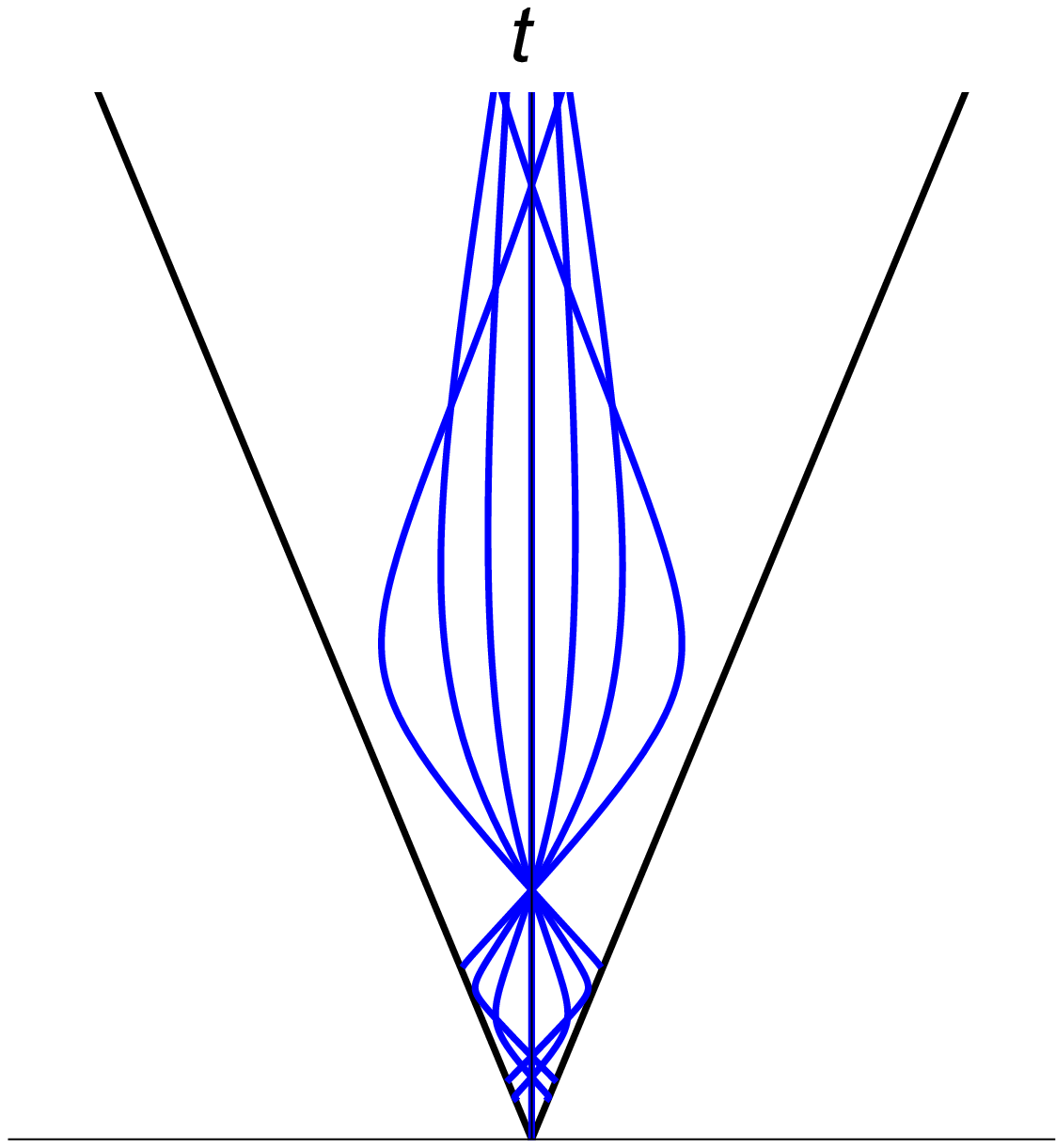}
\label{fig:Jan8}
\caption{\small   Timelike geodesics inside the wedge.}
\end{minipage} 
      \vspace{0pt} \hspace{.1cm}  
\begin{minipage}[t]{0.45\linewidth}
\centering
\includegraphics[scale=.4]{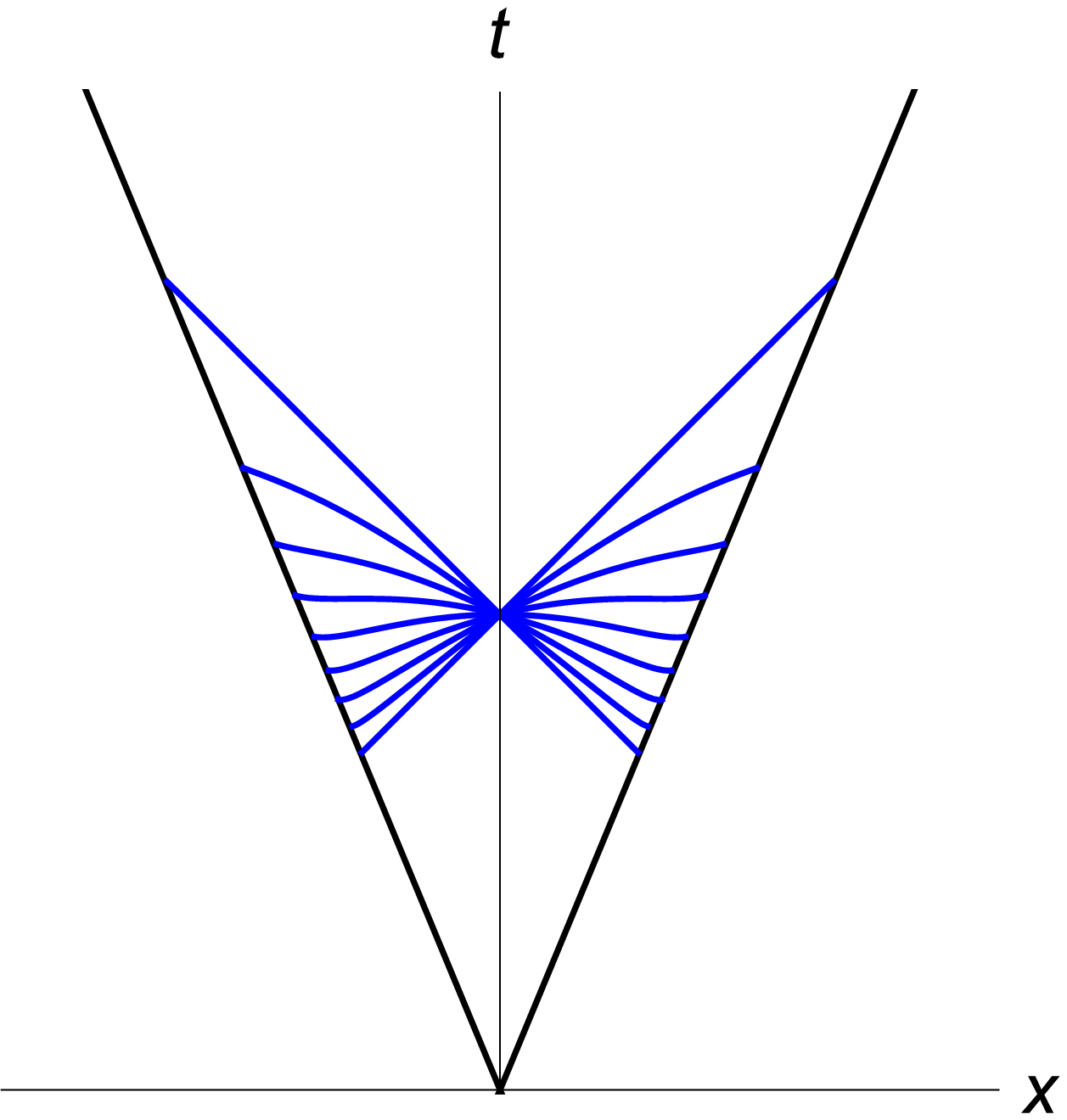}  
\caption{\small Spacelike and null geode- sics inside the wedge.}   
\label{fig:Jan6}  
\end{minipage} 
\end{figure}
\vspace{.cm}

To get a firm grasp of the curved geometry we observe, in Figs. 
3-\ref{fig:Jan6}, that timelike geodesics (inside the 
wedge) tend to converge, and in fact they oscillate back and forth around the 
$t$-axis (corresponding to the spinless Schwarzschild black holes). We are unable to suggest a 
physical interpretation of these geodesics. It might be interesting to investigate how some 
physical processes that may occur close to equilibrium appear when drawn as curves in 
this picture. An example of such a process, which has the double advantage of 
being determined by the parameters of the black hole itself and of having a simple 
analytical form, is accretion of matter from the innermost stable circular orbit of an 
accretion disk. This was described by Bardeen \cite{Bolin}, who showed that an initially 
Schwarzschild black hole of mass $M_i$ will spin up according to the formula 

\begin{equation} a = \sqrt{\frac{2}{3}}\frac{M_i}{M}\left[ 4 - 
\left( \frac{18M_i^2}{M^2} - 2\right)^{1/2} \right] \ , 
\hspace{8mm} M_i \leq M \leq \sqrt{6}M_i 
\ . \end{equation}

\noindent However, the resulting curve starts out spacelike and ends up timelike, so it seems 
fair to say that it is totally insensitive of the Ruppeiner geometry. 

\begin{figure}[h]
        \centerline{ \hbox{
                \epsfig{figure=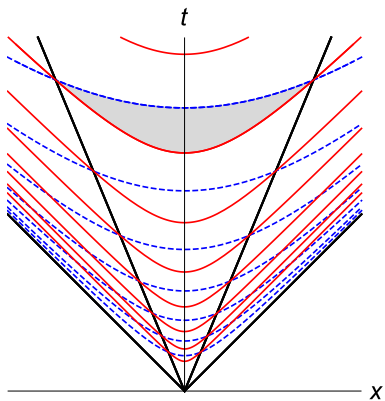,width=50mm}}}
        \caption{\small Contour curves for entropy and mass (dashed). 
        By moving inside the grey area, from near the edge of the wedge (large $a$) 
        towards the centre, one is able to decrease the 
        mass (extract energy) even though the area of the event horizon (the entropy) 
        increases \cite{Christo}, as it must according to Einstein's theory.}
        \label{fig:Jan5}
\end{figure}

The absence of explicit experimental protocols for how to drive a black hole along 
specified curves in state space is deplorable. One obvious target would be the Penrose 
process, briefly alluded to in Fig. \ref{fig:Jan5}, but we are not aware of any suitable 
analytic formulas.

\newpage

{\bf 7. The Kerr-Newman family of black holes}

\

\noindent We have stressed that the scale invariance of Einstein's equations (with a vanishing cosmological constant) 
dictates the form of the entropy function (\ref{skala}), and it is 
this form of the entropy function that dictates the wedge shaped state space associated 
with the outer horizon \cite{Aman2}. For Kerr black holes in five dimensions the picture is 
qualitatively the same as that pertaining to four dimensions, 
but in six or more dimensions black holes of the Myers-Perry variety do not have extreme limits, and 
the wedge shaped state space associated with the outer (only) horizon then fills the entire 
future 
null cone \cite{Aman3, Aman4}. Another example where the wedge fills the entire 
null cone is that of Einstein-Maxwell-dilaton black holes, which again lack an extreme 
limit \cite{Aman5}. In both cases the thermodynamic 
null cone is given by the equation 
$S = 0$, although for the Reissner-Nordstr\"om black holes it is the Ruppeiner metric 
that is the flat one.

The Einstein-Maxwell equations contains a three parameter family of 
black hole equilibrium states. The fundamental relation of 
these Kerr-Newman black holes is

\begin{equation} S= 2M^2 - Q^2 + 2M^2
\sqrt{1 - \frac{Q^2}{M^2} - \frac{J^2}{M^4}} \  \label{Kerr-Na} \end{equation}

\noindent where $J$ is the angular momentum and $Q$ the electric charge of the black hole. 
What is the geometry and global shape of this Gibbs surface? 

It is convenient to rewrite the fundamental relation in terms of dimensionless 
variables, 

\begin{equation} S = M^2f(q,a) \ , \hspace{12mm} 
q = \frac{Q}{M} \ , \hspace{10mm} a = \frac{J}{M^2} \ . \end{equation}

\noindent The form of the function $f$ can be read out from eq. (\ref{Kerr-Na}). The Ruppeiner 
metric is a complicated affair in its defining coordinates. Transforming to the coordinates 
$(\mu , q, a)$, where 

\begin{equation} M = e^{\mu} \ , \end{equation}

\noindent it takes the reasonably attractive form   

\begin{equation} ds^2 = e^{2\mu }\left[ 2(af_{,a}-f)d\mu^2 - 2f_{,q}d\mu dq - 
f_{,qq}dq^2 - 2f_{,qa}dqda - f_{,aa}da^2 \right] \ . \end{equation}

\noindent There is a homothetic Killing vector field 

\begin{equation}  \xi = M\partial_M = \partial_\mu \ , \hspace{8mm} {\cal L}_\xi g = 2g \ . 
\end{equation}

\noindent The boundary of state space occurs for extreme black holes (or when the event 
horizon becomes a degenerate Killing horizon), namely when 

\begin{equation} 
T = 0 \hspace{5mm} \Leftrightarrow \hspace{5mm} 1 = q^2 + a^2 \ . \end{equation}

\noindent The Ruppeiner curvature scalar, and the Ricci tensor in an orthonormal frame, 
are everywhere 
finite in the interior and diverge at the boundary. The actual expressions are complicated 
\cite{Mirza, Ruppeiner4}. 

Thus the Gibbs surface of the Kerr-Newman black holes assumes the shape of a positive cone, 
with the rays defined by the homothetic Killing vector field. This is an interesting shape for 
a state space to assume. The elegance is marred by the fact that the homothetic Killing vector 
field fails to be surface forming. This follows from a simple calculation verifying that 

\begin{equation} \xi_{[a}\nabla_{b}\xi_{c]} \neq 0 \ . \end{equation}

\noindent Other proposals for thermodynamic metrics exist \cite{Quevedo}, and do not suffer 
from this seeming defect \cite{Edward}, but we have not considered them. The homothetic 
Killing vector field is timelike everywhere within the cone corresponding to the outer horizon 
(but this is not so for the homothetic Killing vector field in the Weinhold geometry).

The question whether the inner horizon can be used to extend the Gibbs surface beyond 
the surface of the cone corresponding to the outer horizon
is non-trivial. In the Kerr case we relied on the flat Weinhold 
metric to perform this extension. In the Kerr-Newman case the Weinhold curvature scalar 
remains finite throughout the interior of the cone, but it diverges at the rays $q = \pm 1$ 
corresponding to extreme Reissner-Nordstr\"om black holes. Outside the cone 
it diverges along a two-dimensional surface. None of the two homothetic Killing 
vector fields would be timelike everywhere inside the null cone. 
It is not clear to us how this can 
be handled (if indeed it can be handled).

\

{\bf 8. Discussion}

\

\noindent It seemed appropriate, on this occasion, to review the various uses 
that thermodynamic metrics have been put to---restricting ourselves throughout 
to the Hessian metrics proposed by Weinhold and Ruppeiner. Some of our results 
are (we think) new. We raised and answered the question whether all entropic substances have 
flat thermodynamic metrics. The answer is ``no''. We analytically extended the 
thermodynamic geometry of the Kerr black holes 
to include the entire future 
{\bf null} cone of a two dimensional Lorerentzian space. 
This was achieved by the inclusion of the entropy function for the inner horizon, 
using the Weinhold metric in a key role. 
We went on to clarify in what sense the state space of the Kerr-Newman black holes 
is a cone. It has been known for some time that the scale invariance of Einstein's 
equations implies that wedge shaped state spaces in the interior of the thermodynamic 
null cone arise for all sorts of black hole families, but these analyses were concerned 
with the entropy function defined by the outer horizon only. 

There are many things we would like to know, but don't. First of all we 
have not analyzed what restrictions on the function $f$, in entropy functions of 
the form given in eq. (\ref{skala}), make the extension of the wedge to include 
the entire light cone possible. It does not seem to be straightforward in the 
Reissner-Nordstr\"om case, once we insist that the extension be related to the entropy 
function of the inner horizon. Moreover, and crucially, a cosmological constant 
introduces a new scale and complicates matters. This also is known \cite{Aman1}, 
but not in anything like the detail that we would like to see. The entropy 
function will still be approximately a generalized homogeneous function close to 
the origin, so some features of our analysis will survive. 

We have certainly not reached the point where we can tell from the shape of the Gibbs 
surface that we are dealing with black hole thermodynamics. 

\subsection*{Acknowledgements}

We thank John Ward, Daniel Grumiller, George Ruppeiner, Laszlo Gergely, Roberto Emparan among others for many discussions over the years.

\end{document}